\documentclass[11pt]{article}
\usepackage{epsf}

\begin{document}

%theorem environments:

\newtheorem{theorem}{Theorem}[section]
\newtheorem{itlemma}{Lemma}[section]
\newtheorem{itdefinition}{Definition}[section]
\newtheorem{itexample}{Example}
\newtheorem{itclaim}{Claim}[section]
\newtheorem{itproposition}{Proposition}[section]
\newtheorem{itremark}{Remark}[section]
\newtheorem{itcorollary}{Corollary}[section]

%the same environments without italics:
\newenvironment{example}{\begin{itexample}\rm}{\end{itexample}}
\newenvironment{definition}{\begin{itdefinition}\rm}{\end{itdefinition}}
\newenvironment{lemma}{\begin{itlemma}\rm}{\end{itlemma}}
\newenvironment{corollary}{\begin{itcorollary}\rm}{\end{itcorollary}}
\newenvironment{claim}{\begin{itclaim}\rm}{\end{itclaim}}
\newenvironment{proposition}{\begin{itproposition}\rm}{\end{itproposition}}
\newenvironment{remark}{\begin{itremark}\rm}{\end{itremark}}

\newcommand{\qed}{\hfill \halmos} %put \qed at right margin
\newcommand{\mybox}{\hfill $\Box$} %put \qed at right margin (white square)

\newcommand{\comment}[1]{}
\newcommand{\halmos}{\rule{1ex}{1.4ex}}
%alternative to \qed (see below)
\newenvironment{proof}{\noindent {\em Proof}.\ }{\hspace*{\fill}$\halmos$
\medskip}

%mathematical symbols:
\def\vbar{\mathchoice{\vrule height6.3ptdepth-.5ptwidth.8pt\kern-.8pt}
   {\vrule height6.3ptdepth-.5ptwidth.8pt\kern-.8pt}
   {\vrule height4.1ptdepth-.35ptwidth.6pt\kern-.6pt}
   {\vrule height3.1ptdepth-.25ptwidth.5pt\kern-.5pt}}
\def\fudge{\mathchoice{}{}{\mkern.5mu}{\mkern.8mu}}
\def\bbc#1#2{{\rm \mkern#2mu\vbar\mkern-#2mu#1}}
\def\bbb#1{{\rm I\mkern-3.5mu #1}}
\def\bba#1#2{{\rm #1\mkern-#2mu\fudge #1}}
\def\bb#1{{\count4=`#1 \advance\count4by-64 \ifcase\count4\or\bba A{11.5}\or
   \bbb B\or\bbc C{5}\or\bbb D\or\bbb E\or\bbb F \or\bbc G{5}\or\bbb H\or
   \bbb I\or\bbc J{3}\or\bbb K\or\bbb L \or\bbb M\or\bbb N\or\bbc O{5} \or
   \bbb P\or\bbc Q{5}\or\bbb R\or\bbc S{4.2}\or\bba T{10.5}\or\bbc U{5}\or
%   \bbb P\or\bbc Q{5}\or\bbb R\or\bba S{8}\or\bba T{10.5}\or\bbc U{5}\or
   \bba V{12}\or\bba W{16.5}\or\bba X{11}\or\bba Y{11.7}\or\bba Z{7.5}\fi}}

\def\Q{{\bb Q}}                         %rationals
\def\N{{\bb N}}                         %naturals
\def\R{{\bb R}}                         %real%
\def\I{{\bb Z}}                         %integers
\def\B{{\bb B}}                         %in {0,1}*

\def\rmtr{{\rm tr}}

%%%asymptotics%%%%%%%%
\def\nasymptotic{{_{\stackrel{\displaystyle\longrightarrow}
{N\rightarrow\infty}}\,\, }} %N  goe to infinity, display sty.
\def\masymptotic{{_{\stackrel{\displaystyle\longrightarrow}
{M\rightarrow\infty}}\,\, }} %M  goe to infinity, display sty.
\def\wasymptotic{{_{\stackrel{\displaystyle\longrightarrow}
{w\rightarrow\infty}}\,\, }} %w goe to infinity, display sty.

\def\asymptext{\raisebox{.6ex}{${_{\stackrel{\displaystyle\longrightarro
w}{x\rightarrow\pm\infty}}\,\, }$}} %x goes to plus minus infinity, within
\def\epsilim{{_{\textstyle{\rm lim}}\atop_{\epsilon\rightarrow 0+}\,\, }}
%epsilon goes to zero, display

%%%%%%equations and eq arrays %%%%%%
\def\beqra{\begin{eqnarray}} \def\eeqra{\end{eqnarray}}
\def\beqast{\begin{eqnarray*}} \def\eeqast{\end{eqnarray*}}
\def\beq{\begin{equation}}      \def\eeq{\end{equation}}
\def\be{\begin{enumerate}}   \def\ee{\end{enumerate}}

%%%%%%%greek%%%%%%%%%
\def\bet{\beta}
\def\gam{\gamma}
\def\Gam{\Gamma}
\def\la{\lambda}
\def\eps{\epsilon}
\def\La{\Lambda}
\def\si{\sigma}
\def\Si{\Sigma}
\def\al{\alpha}
\def\Tha{\Theta}
\def\tha{\theta}
\def\vphi{\varphi}
\def\del{\delta}
\def\Del{\Delta}
\def\ab{\alpha\beta}
\def\om{\omega}
\def\Om{\Omega}
\def\mn{\mu\nu}
\def\mun{^{\mu}{}_{\nu}}
\def\kap{\kappa}
\def\rsi{\rho\sigma}
\def\beal{\beta\alpha}

%%%%%%misc%%%%%%5
\def\til{\tilde}
\def\rta{\rightarrow}
\def\eqv{\equiv}
\def\nab{\nabla}
\def\pa{\partial}
\def\sit{\tilde\sigma}
\def\ul{\underline}
\def\indt{\parindent2.5em}
\def\nd{\noindent}
\def\var{{1\over 2\si^2}}
\def\ivar{\left(2\pi\si^2\right)}
\def\iivar{\left({1\over 2\pi\si^2}\right)}
%%%%%%%%%%%%% caligraphic%%%%%%%
\def\caa{{\cal A}}
\def\cb{{\cal B}}
\def\cac{{\cal C}}
\def\cd{{\cal D}}
\def\ce{{\cal E}}
\def\cf{{\cal F}}
\def\cg{{\cal G}}
\def\ch{{\cal H}}
\def\ci{{\cal I}}
\def\cj{{\cal{J}}}
\def\ck{{\cal K}}
\def\cl{{\cal L}}
\def\cm{{\cal M}}
\def\cn{{\cal N}}
\def\cO{{\cal O}}
\def\cp{{\cal P}}
\def\cq{{\cal Q}}
\def\car{{\cal R}}
\def\cs{{\cal S}}
\def\ct{{\cal{T}}}
\def\cu{{\cal{U}}}
\def\cv{{\cal{V}}}
\def\cw{{\cal{W}}}
\def\cx{{\cal{X}}}
\def\cy{{\cal{Y}}}
\def\cz{{\cal{Z}}}

%
%=============================================
\vspace*{-0.5in}
\begin{center}
{\large\bf
Random matrix theory for the analysis of the performance of an analog computer: a scaling theory}
\end{center}
\vspace{-.2in}
\begin{center}
{\bf Asa Ben-Hur$^{a}$, Joshua Feinberg$^{b,c}$, Shmuel Fishman$^{c,d}$}\\
{\bf \& Hava T. Siegelmann$^{e}$}
\end{center}
%\vskip 1mm
\begin{center}
$^{a)}${Biochemistry Department,}\\ 
{Stanford University, CA 94305, USA}\\
$^{b)}${Physics Department,}\\
{University of Haifa at Oranim, Tivon 36006, Israel}\\
$^{c)}${Physics Department,}\\
{Technion, Israel Institute of Technology, Haifa 32000, Israel}\\
%\vskip 2mm
$^{d)}${Institute for Theoretical Physics}\\
{University of California\\ Santa Barbara, CA 93106, USA}\\
$^{e)}${Bio-Computation Laboratory,}\\
{Department of Computer Science, University of Massachusetts,}\\ 
{Amherst MA 01003}\\
%\vskip 2mm
\end{center}
%\vskip 3mm

%\date{October 2001}
%\maketitle
%\vskip 3mm
\begin{minipage}{6.1in}
{\abstract
The phase space flow of a dynamical system, leading to the
solution of Linear Programming (LP) problems, is explored as
an example of complexity analysis in an analog computation
framework. In this framework, computation by physical
devices and natural systems, evolving in continuous phase
space and time (in contrast to the digital computer where
these are discrete), is explored.  A Gaussian ensemble of LP
problems is studied. The convergence time of a flow to the
fixed point representing the optimal solution, is computed.
The cumulative distribution function of the convergence time is
calculated in the framework of Random Matrix Theory (RMT) in
the asymptotic limit of large problem size. It is found to
be a scaling function, of the form obtained in the theories
of critical phenomena and Anderson localization.  It
demonstrates a correspondence between problems of Computer
Science and Physics.}
\end{minipage}

\vspace{10pt}
%{\bf Keywords:} Theory of Analog Computation, Dynamical Systems, Linear
%Programming, Scaling, Random Matrix Theory.\\
%e-mail addresses: asa@barnhilltechnologies.com,~
%joshua@physics.technion.ac.il,\\ fishman@physics.technion.ac.il,~
%hava@mit.edu

%\maketitle
%=================================================

An analog computer is a physical device that performs
computation, evolving in continuous time and phase space;
its evolution in phase space can be modeled by dynamical 
systems (DS)  \cite{ott}, the way classical systems such as
particles moving in a potential (or electric circuits), are
modeled. This description makes a large set of analytical tools 
and physical intuition,
developed for dynamical systems, applicable to the analysis of analog 
computers.
In contrast, the evolution of a digital computer is described by a
dynamical system, discrete both in its phase space
and in time. The most relevant examples of analog computers are
VLSI devices implementing neural networks
\cite{Hertznn-optimwang}, or neuromorphic systems
\cite{mead}, whose structure is directly motivated by the
workings of the brain. Various processes taking place in
living cells can be considered as analog
computation \cite{bio}. Dynamical systems (described by
ordinary differential equations) are also used
to solve computational problems \cite{brockett,faybusovich,helmke-moore}. The
standard theory of computation and computational complexity
\cite{Papadimitriou} deals with computation in discrete time
and phase space, and is inadequate for the description of
such systems. For the analysis of computation by analog
devices a theory that is valid in continuous time and phase
space is required.  Since the systems in question are
physical systems, the computation time is the time required
for a system to reach the vicinity of an attractor (a stable
fixed point in the present work) combined with the time
required to verify that it indeed reached this vicinity.
This time is the elapsed time measured by a clock, contrary
to standard computation theory, where it is the number of
steps.

In the exploration of physical systems, it is sometimes much easier to study
statistical ensembles of systems, estimating their typical behavior using
statistical methods \cite{rmt,nucl,chaos}.  Ensembles of systems modeling the
dynamics of populations were studied as well \cite{may,other}. The statistical
theories describe many general features of the problems that are inverstigated,
but specific systems require special attention \cite{nucl,other}. In this
letter a statistical theory is used to calculate the computational complexity
of a standard representative problem, namely Linear Programming (LP), as 
solved by a DS.  A detailed version was published in the computer 
science literature \cite{bffs}.

In two recent papers we have proposed a framework for
computing with DS that converge exponentially to fixed
points \cite{dds}. For such systems it is natural to
consider the {\em attracting fixed point as the output}. The
input can be modeled in various ways. One possible choice is
the initial condition. This is appropriate when the aim of
the computation is to decide to which attractor out of many
possible ones the system flows \cite{SF}. Here, as in
\cite{dds}, the parameters on which the system of DS depends
(e.g., the parameters appearing in the vector field $F$ in
(\ref{contsys})) are the input.

The basic entity of the computational model is a dynamical
system \cite{ott}, that may be defined by a set of
Ordinary Differential Equations (ODEs)
\beq\label{ode}
\frac{dx}{dt}=F(x),
\label{contsys}
\eeq
where $x$ is an $n$-dimensional vector, and $F$ is an
$n$-dimensional smooth vector field, which converges
exponentially to a fixed point. Eq. (\ref{contsys})
solves a computational problem as follows: Given an instance
of the problem, the parameters of the vector field $F$ are
set, and it is started from some pre-determined initial
condition. The result of the computation is then deduced
from the fixed point that the system approaches.

In our model we assume we have a physical implementation of the flow 
equation (\ref{contsys}). Thus, the vector field $F$ need not be computed, 
and the computation time is determined by the convergence time to the 
attractive fixed point. In other words, the time of flow to the vicinity
of the attractor is a good measure of complexity, namely the computational 
effort, for the class of continuous dynamical systems introduced above 
\cite{dds}.

In this letter we will consider real inputs, as 
the ones found in physical experiments, and that are studied in the BSS
model \cite{BCSS}. For computational models defined on the
real numbers, worst case behavior, that is traditionally
studied in computer science, can be ill defined and lead to
infinite computation times, in particular, for some methods
for solving LP \cite{BCSS, traub}. Therefore, we compute the
distribution of computation times for a probabilistic model
of LP instances with Gaussian distribution of the data like
in \cite{lpsmaleTodd-models,shamir}. Ill-defined instances 
constitute a set of zero measure in our probability ensemble, and need 
not be concerned about.

The computational complexity of the method 
presented here is $\cO(n\log n)$, compared to
$\cO(n^{3.5} \log n)$ found for standard interior point
methods \cite{Ye-book}. The basic
reason is that for standard methods (such as interior point methods), the 
major component of the complexity of each
iteration is $\cO(n^3)$ due to matrix decomposition and inversion of the 
constraint matrix, while here, because of its analog nature, the system 
just flows according to its equations of motion (which need not be computed).

Since we consider the evolution of a {\em vector filed}, our model is
inherently parallel. Therefore, to make the analog vs. digital comparison
entirely fair, we should compare the complexity of our method to that of the 
best parallel algorithm. The latter can reduce the $\cO(n^3)$ time needed for 
matrix decomposition/inversion to polylogarithmic time (for well-posed 
problems), at the cost of $\cO(n^{2.5})$ processors \cite{reif}, while our 
system of equations (\ref{contsys}) uses only $\cO(n)$ variables.

Linear programming is a P-complete problem \cite{Papadimitriou}, i.e.
it is representative of all problems that can be solved in polynomial time.
The {\it standard form} of LP is
to find \begin{equation}
\label{standard}
\max \{ c^{T}x~:~ x\in \R ^{n},
  A x =  b,x \geq 0  \}
\end{equation}
where $c \in \R^n, b \in \R^m,
A \in \R^{m \times n}$
and $m\leq n$.
The set generated by the constraints in (\ref{standard}) is a polyheder.
If a bounded optimal solution exists, it is obtained at one of its vertices.
The vector defining this optimal vertex can be decomposed
(in an appropriate basis) in the form
$x=(x_{{\cal N}},x_{{\cal B}})$
where $x_{\cal N} = 0$ is an $n-m$ component
vector, while $x_{\cal B} =  B^{-1}  b \geq  0$
is an $m$ component vector, and $B$ is the $m \times m$ matrix whose
columns are
the columns of $A$ with indices identical to the ones of $x_{\cal B}$.
Similarly, we decompose $A=(N,B)$.

A flow of the form (\ref{contsys}) converging to the optimal vertex,
introduced by Faybusovich \cite{faybusovich} will be studied here.
Its vector field $F$ is a projection of the gradient of the
cost function $c^T x$ onto the constraint set, relative to a
Riemannian metric which enforces the positivity constraints $x\geq 0$
\cite{faybusovich}. It is given by
\begin{equation} \label{field}
	F(x)=[X - X A^{T}
		(A X A^{T})^{-1} A X]\: c\; ,
\end{equation}
where $X$ is the diagonal matrix $\mbox{Diag}(x_1 \dots x_n)$.
The $nm + n$ entries of $A$ and $c$, namely, the parameters of the 
vector field $F$, constitute the input; as in other models of 
computation, we ignore the time it takes to ``load'' the input, since this 
step does not reflect the complexity of the computation being performed, 
either in analog or digital computation.
It was shown in \cite{toda} that the flow equations 
given by
(\ref{contsys}) and (\ref{field}) are, in fact, part of a system of
Hamiltonian equations of motion of a completely integrable system of a
Toda type. Therefore, like the Toda system, it is integrable with the
formal solution \cite{faybusovich}
\begin{equation}
\label{solution}
x_i(t) = x_i(0) \exp \left( -\Delta_i t +
\sum_{j=1}^{m} \alpha_{ji} \log \frac{x_{j+n-m}(t)}{x_{j+n-m}(0)} \right)
\end{equation}
($i = 1,\ldots ,n-m$), that describes the time evolution of the $n-m$
independent variables $x_{\cal N}(t)$, in terms of the variables
$x_{\cal B}(t)$. In (\ref{solution})
$x_i(0)$ and $x_{j+n-m}(0)$ are components of the initial
condition, $x_{j+n-m}(t)$ are the $x_{\cb}$ components of the solution,
$\alpha_{ji}= - (B^{-1} N)_{ji}$
is an $m \times (n-m)$ matrix, while
\begin{equation}\label{deltas}
\Delta_i = -c_i  - \sum_{j=1}^{m} c_j \alpha_{ji}\,.
\end{equation}
For the decomposition
$x=(x_{{\cal N}},x_{{\cal B}})$
used for the optimal vertex $\Delta_i \geq 0~~i=1,\ldots,n-m\,,$
and $x_{\cal N}(t)$ converges to 0, while
$x_{\cal B}(t)$ converges to $x^*=B^{-1}b$.
Note that the analytical solution is only a {\em formal} one, and does not
provide an answer
to the LP instance, since the $\Delta_i$ depend on the partition of $A$, and
only relative to a partition corresponding to a
maximum vertex are all the $\Delta_i$ positive.

The second term in (\ref{solution}),
when it is positive, is a kind of ``barrier'':
$\Delta_{i}t$
must be larger than the barrier before $x_i$ can decrease to zero.
In the following we ignore the contribution of the initial condition
and denote the value of this term in the infinite time limit by
\begin{equation}
\label{barrier}
\beta_i = \sum_{j=1}^m \alpha_{ji} \log x_{j+n-m}^*.
\end{equation}
Note that although one of the $x_j^*$ may vanish, in the probabilistic
ensemble studied here such an event is of measure zero and therefore
should not be considered.
In order for $x(t)$ to be close to the maximum vertex we must have
$x_i(t) < \epsilon$ for $i=1,\ldots,n-m$ for some small positive
$\epsilon$, namely
%\begin{equation}\label{less_than_eps}
$\exp (- \Delta_i t + \beta_i) < \epsilon ~,~~ \mbox{for}~ i =
1,\ldots,n-m.$
%\end{equation}
Therefore we consider
\begin{equation}\label{T}
T = \max_{i} \left( \frac{\beta_i}{\Delta_i} +
    \frac{|\log \epsilon |}{\Delta_i} \right)~,
\end{equation}
as the computation time.
We denote
\begin{equation}\label{Deltamin}
\Delta_{\min} = \min_i \Delta_i,~~~\beta_{\max} = \max_i \beta_i \;.
\end{equation}
The $\Delta_i$ can be arbitrarily small when the inputs are real numbers,
but in the probabilistic model,
``bad'' instances are rare as is clear from (\ref{scaling.delta}).

The ensemble we analyze consists of
LP problems in which the components of $(A,b,c)$ are independent
identically distributed (i.i.d.) random variables taken from
the standard Gaussian distribution
with 0 mean and unit variance.
With the introduction of a probabilistic model of LP instances,
$\Delta_{\min},\, \beta_{\max}$ and $T$ become
random variables.
Since the expression for $\Delta_i$, equation (\ref{deltas}),
is independent of $b$, its distribution is independent of $b$.
For a given realization of $A$ and $c$, with a partition of $A$
into $(N,B)$ such that $\Delta_i \geq 0$, there exists a vector $b$ such
that the resulting polyheder has a bounded optimal solution.
Since $b$ in our probabilistic model is independent of $A$ we obtain:
\newline
${\cal P}(\Delta_{\min} < \Delta | \Delta_{\min} > 0,
\mbox{LP instance has a bounded maximum vertex}) =
{\cal P}(\Delta_{\min} < \Delta | \Delta_{\min} > 0)$.

We wish to compute the probability distribution of $\Delta_{\min}$
for instances with a bounded solution, when $\Delta_{\min} > 0$, denoted by
${\cal P}(\Delta_{min} > \Delta | \Delta_{min} > 0)$.
It turns out that it is much easier to analytically calculate the probability
distribution of $\Delta_{min}$ for a given partition of the matrix $A$.
In the probabilistic model we defined,
${\cal P}(\Delta_{\min} > \Delta | \Delta_{\min} > 0)$ is proportional
to the probability that $\Delta_{\min} > \Delta$ for a fixed partition
(\ref{lemma3.2}).
Let the index 1 stand for the partition where $B$ is taken from the last
$m$ columns of $A$.
In \cite{bffs} we proved, using the symmetry resulting from the identity of the
Gaussian variables, that
\beq\label{lemma3.2}
{\cal P}(\Delta_{\min} > \Delta | \Delta_{\min} > 0) =
\frac{{\cal P}(\Delta_{min1} > \Delta)}{{\cal P}(\Delta_{min1} > 0)}
\eeq
for  $\Delta > 0$, where
$
\Delta_{min1} = \min \{\Delta_i ~|~ \Delta_i
\mbox{~are computed relative to the partition 1} \}\,$
%\begin{equation}
%\Delta_{min1} = \min \{\Delta_i ~|~ \Delta_i
%\mbox{~are computed relative to the partition 1} \}\,.
%\end{equation}
and ${\cal P}(\Delta_{min1}>0) = 1/2^{n-m} \;$.
%\begin{equation}\label{Ppositive}
%{\cal P}(\Delta_{min1}>0) = 1/2^{n-m} \;.
%\end{equation}

Integrating over the Gaussian variables of the ensemble, the probability
${\cal P}(\Delta_{min1} > \Delta)$
was computed in \cite{bffs} for a specific partition of $A$ in the large
$(n,m)$ asymptotic limit, making use of methods of random matrix theory.
Given ${\cal P}(\Delta_{min1} > \Delta)$, then
${\cal P}(\Delta_{\min} >\Delta | \Delta_{\min} > 0)$
is obtained with the
help of (\ref{lemma3.2}).
In the large $(n,m)$ limit the probability ${\cal P}(\Delta_{min} < \Delta|
\Delta_{min} > 0) \equiv {\cal F}^{(n,m)}(\Delta)$
is of the scaling form
\begin{equation} \label{scaling.delta}
{\cal F}^{(n,m)}(\Delta)=1-e^{x_\Delta^2}\,{\rm erfc}(x_\Delta)\ \equiv {\cal
F}(x_\Delta) .
\end{equation}
%where
%\begin{equation}\label{scalingvariable}
%x_\Delta = \eta_\Delta (n,m) \Delta\,,
%\end{equation}
with the scaling variable $x_\Delta (n,m) =  {1\over\sqrt{\pi}}\left({n\over
m}-1\right)\,\sqrt{m} \Delta\, .$
The scaling function ${\cal F}$ contains {\em all} asymptotic information on $\Delta$.
The distribution ${\cal F}(x_\Delta)$ is very wide and
does not have a mean.
Also the average of $1/x_\Delta$ diverges.

In order to the demonstrate this result numerically, we generated
LP instances where $(A,b,c)$
are random Gaussian variables and
solved for them the LP problem with the IMSL C library.
We obtained an estimate of
${\cal F}^{(n,m)}(\Delta)={\cal P}(\Delta_{min} < \Delta |
\Delta_{min} > 0)$, and of the corresponding cumulative distribution
functions of the barrier $\beta_{\max}$ and $T$.
In Fig. 1
the numerical results are compared with the analytical formula
(\ref{scaling.delta}).

\begin{figure}
\epsfxsize=8cm
\centerline{\epsffile[60 200 550 610]{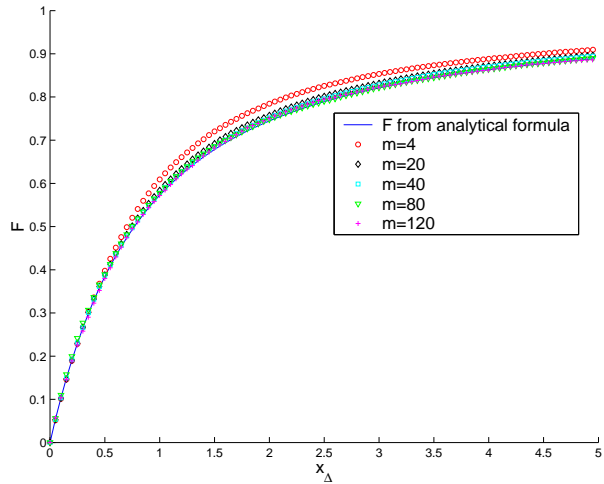}}
\caption{\label{graph-xdelta}
\ensuremath{ {\cal F}(x_\Delta) }
is plotted against the variable
\ensuremath{ x_{\Delta}},
for \ensuremath{n=2m}. There is very good agreement with the
analytical results, improving as $m$ increases, as expected for an asymptotic result.}
\end{figure}

The existence of scaling functions like (\ref{scaling.delta}) for the
barrier
$\beta_{max}$, that is the maximum of the $\beta_i$ defined by (\ref{barrier})
and for  $T$ defined by (\ref{T}) was verified numerically (see
Fig. 2 for $T$).
In particular for fixed $m/n$, we found that
\begin{equation} \label{scaling.beta}
{\cal P}(1/\beta_{max}~<~1/\beta) \equiv {\cal
F}^{(n,m)}_{1/\beta_{max}}(1/\beta)= {\cal F}_{1/{\beta_{max}}}(x_\beta)
\end{equation}
and
\begin{equation} \label{scaling.T}
{\cal P}(1/T~<~1/t) \equiv {\cal
F}^{(n,m)}_{1/T}(1/t)= {\cal F}_{1/T}(x_T).
\end{equation}
The scaling variables are $x_\beta \sim m/\beta$ and $x_T \sim m \log m/t$.
\begin{figure}
\epsfxsize=8cm
\centerline{\epsffile[60 200 550 610]{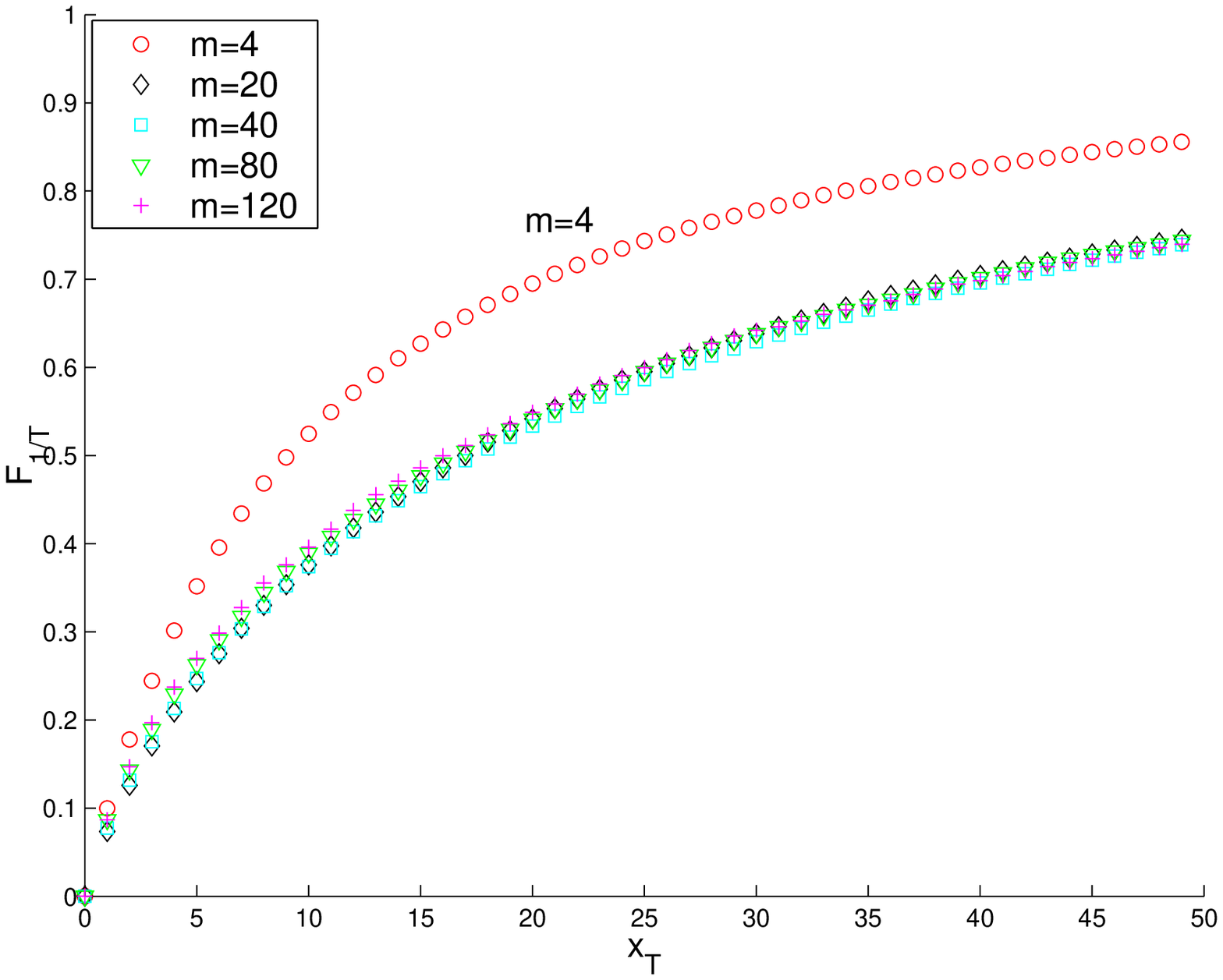}}
\caption{\label{graph-T}
\ensuremath{ {\cal F}_{1/T}(x_{T}) } as a function of the variable
\ensuremath{x_{T} = m \log m /t } for the same instances as Fig. 1.
}
\end{figure}
The scaling functions (\ref{scaling.delta}),  (\ref{scaling.beta}) and
(\ref{scaling.T})
imply the asymptotic behavior
\begin{eqnarray}\label{om}
 1/\Delta_{\min} \sim \sqrt{m}, ~~~~
\beta_{\max} \sim m ,~~~
T \sim m \log m
\end{eqnarray}
with ``high probability'' \cite{bffs}.

In this letter we computed the problem size dependence of
the distributions of quantities that govern the convergence
of a DS that solves the LP problem \cite{faybusovich}. To
the best of our knowledge, this is the first time such
distributions are computed. In particular, knowledge of the
distribution functions enables to obtain the ``high
probability'' behavior (\ref{om}), and the moments (if these
exist). The main result of the present work is that the
distribution functions of the convergence rate,
$\Delta_{min}$, the barrier $\beta_{max}$ and the
computation time $T$ are scaling functions; i.e., in the
asymptotic limit of large $(n,m)$, each depends on the
problem size only through a scaling variable. In other words
these are not arbitrary functions of the three variables, but
each is a function only of one variable, $x_\Delta$, $x_\beta$
or $x_T$. The distribution function of $\Delta_{min}$ was
calculated analytically, and the result was verified
numerically. The scaling functions, even if known only
numerically, can be useful for the understanding of the
behavior for large values of $(n,m)$ that are beyond the
limits of numerical simulations.

In this letter the distribution functions of various
quantities that characterize the computational complexity,
were found to be scaling functions in the large $(n,m)$
limit. This is analogous to the situation found for the
central limit theorem, for critical phenomena \cite{wilson}
and for Anderson localization \cite{anderson}, in spite of the very different 
nature of these problems.
It is demonstrated here how for the implementation
of the LP problem on a physical device, methods used in
theoretical physics enable to describe the distribution
of computation times in a simple and physically transparent form.
Based on our experience with certain universality properties of
rectangular and chiral random matrix models \cite{universality},
we expect some universality for computational problems, that
should be explored. The obvious questions are: Is the
Gaussian nature of the ensemble unimportant in analogy with
\cite{universality}? 
Are there universality classes \cite{wilson}
of analog computational problems, and if they exist, what are they? 
Are these analogous to the classification of \cite{other}?
We believe it can be instructive to explore computational problems using 
methodologies of theoretical physics as was
demonstrated here for linear programming.

%%%%%%%%%%%%%%%%%%%%%%%%%%%%%%%%%%%%%%%%%%%%%%%%%%%%%%%%%%%%%%%%%%%%%%%%%%%%%%
%%%%%%%%%%%%%%%%%%%%%%%%%%%%%%%%%%%%%%%%%%%%%%%%%%%%%%%%%%%%%%%%%%%%%%%%%%%%%%

Acknowledgements:
It is our great pleasure to thank Arkadi Nemirovski,  Eduardo Sontag and Ofer
Zeitouni for stimulating and informative discussions.
We thank a referee for bringing \cite{may,other} to our attention.
This research was supported
in part by the US-Israel Binational Science Foundation
(BSF), by the Israeli Science Foundation, by the US National Science
Foundation under Grant No. PHY99-07949 and by the Minerva Center of
Nonlinear Physics of Complex Systems.

\end{document}